\pdfoutput=1
\documentclass[]{aa} 

\usepackage{graphicx}
\usepackage{txfonts}
\usepackage{hyperref}

\def\thetaS{\mbox{\boldmath$\theta_{\rm S}$}}
\def\thetaI{\mbox{\boldmath$\theta_{\rm I}$}}

\def\angle0{\mbox{\boldmath$\theta_{\rm planet}$}}
\def\starangle{\mbox{\boldmath$\theta_{*}$}}

\def\Deltabeta{\mbox{\boldmath$\Deltabeta$}}

\begin{document} 

\title{Refraction in exoplanet atmospheres}
\subtitle{Photometric signatures, implications for transmission spectroscopy, and search in \textit{Kepler} data}
\author{D.~Alp
  \inst{1}
  \and
  B.-O.~Demory\inst{2}
}
\authorrunning{D.~Alp \& B.-O.~Demory}
\institute{Department of Physics, KTH Royal Institute of Technology, 
  The Oskar Klein Centre, AlbaNova, SE-106 91 Stockholm, Sweden\\
  \email{dalp@kth.se}
  \and
  Center for Space and Habitability, University of Bern, Sidlerstrasse
  5, 3012 Bern, Switzerland\\
  \email{brice.demory@csh.unibe.ch}
}
\date{Received 30 June 2017 / Accepted 27 October 2017}

\abstract
{Refraction deflects photons that pass through atmospheres, which
  affects transit light curves. Refraction thus provides an avenue to
  probe physical properties of exoplanet atmospheres and to constrain
  the presence of clouds and hazes. In addition, an effective surface
  can be imposed by refraction, thereby limiting the pressure levels
  probed by transmission spectroscopy.}
{The main objective of the paper is to model the effects of refraction
  on photometric light curves for realistic planets and to explore the
  dependencies on atmospheric physical parameters. We also explore
  under which circumstances transmission spectra are significantly
  affected by refraction. Finally, we search for refraction signatures
  in photometric residuals in \textit{Kepler} data.}
{We use the model of \citet{hui02} to compute deflection angles and
  refraction transit light curves, allowing us to explore the
  parameter space of atmospheric properties. The observational search
  is performed by stacking large samples of transit light curves from
  \textit{Kepler}.}
{We find that out-of-transit refraction shoulders are the most easily
  observable features, which can reach peak amplitudes of $\sim$10
  parts per million (ppm) for planets around Sun-like stars. More
  typical amplitudes are a few ppm or less for Jovians and at the
  sub-ppm level for super-Earths. In-transit, ingress, and egress
  refraction features are challenging to detect because of the short
  timescales and degeneracies with other transit model
  parameters. Interestingly, the signal-to-noise ratio of any
  refraction residuals for planets orbiting Sun-like hosts are
  expected to be similar for planets orbiting red dwarfs. We also find
  that the maximum depth probed by transmission spectroscopy is not
  limited by refraction for weakly lensing planets, but that the
  incidence of refraction can vary significantly for strongly lensing
  planets. We find no signs of refraction features in the stacked
  \textit{Kepler} light curves, which is in agreement with our model
  predictions.}
{}

\maketitle

\section{Introduction}\label{sec:introduction}
The transit method is one of the most successful at finding
exoplanets, with thousands of detections by the \textit{Kepler}
mission~\citep{koch10} alone. Transit light curves are typically
modelled as planets occulting host stars with non-uniform brightness
profiles~\citep{mandel02}. Light curves with higher photometric
precision reveal additional effects, such as thermal emission from
planets~\citep[e.g.][]{charbonneau05, demory12} and reflected host
star light~\citep[e.g.][]{sudarsky00, demory14}. Searches for rings
and moons have also been carried out~\citep{hippke15, kipping15,
  heller17}.  Further development of light-curve models is therefore
desirable in order to account for physical effects that were
previously not included, especially on the verge of the commissioning
of new facilities that will significantly improve the current
photometric capabilities.

Refraction in exoplanet atmospheres, or `atmospheric lensing', was
first discussed by \citet{seager00} and \citet{hubbard01}, who
concluded that refraction is weak at low pressures. \citet{hui02}
presented a model and made the first detailed study of the effects of
refraction and oblateness on transit light curves. \citet{sidis10}
studied refraction in the extreme case of a transparent planet and
showed that dips in light curves can be caused by refractive
transparent planets. \citet{betremieux15} provide analytic expressions
for the column density and deflection angle, and make comparisons with
a ray-tracing algorithm. Refraction in exoplanet atmospheres have
received increased attention over the past few years and several
authors have emphasised that refraction can set the effective planet
radius~\citep{garcia_munoz12, betremieux13, betremieux14}. The
implications of the effective surface depth due to refraction on
transmission spectroscopy have also been studied~\citep{betremieux16,
  betremieux17}. Several authors have studied how transmission spectra
depend on transit phase and the possibility of using refraction to
probe the atmospheric composition at different
altitudes~\citep{sidis10, garcia_munoz12, misra14}. Refraction
signatures in photometric light curves have been proposed by
\citet{misra14b} to serve as an efficient method of identifying clear
skies. No attempt has yet been made to detect refraction signals in
observational data despite the recent theoretical developments.

This paper studies the effects of refraction on transit light curves
and how refraction signals depend on physical properties of exoplanet
atmospheres. We compute the expected signal strengths and
characteristics for a sample of planets. We then discuss the effects
of refraction on planetary transmission spectra. We finally survey the
complete \textit{Kepler} primary mission dataset to search for
refraction features.

The paper is organised as follows. We describe the model and suite of
test planets in Sect.~\ref{sec:modelling}, and present model
predictions in Sect.~\ref{sec:mod_results}. Observations and data
reduction are detailed in Sect.~\ref{sec:observations}, and
observational results are presented in Sect.~\ref{sec:obs_results}.
We discuss our findings in Sect.~\ref{sec:discussion}, and summarise
our conclusions in Sect.~\ref{sec:conclusions}.

\section{Model description}\label{sec:modelling}
We use the following conventions and definitions: residuals are
defined as $\Delta F = F_\mathrm{R}-F_\mathrm{M}$, where
$F_\mathrm{R}$ is the normalised observed or predicted model flux and
$F_\mathrm{M}$ is the normalised flux of a fitted model. All
uncertainties are one standard deviation. We parametrise the stellar
surface brightness as
$I = 1 - \gamma_1\,(1-\nu) - \gamma_2\,(1-\nu)^2$, where $\gamma_1$
and $\gamma_2$ are the limb-darkening coefficients (LDCs) and
$\nu = \sqrt{1-r^2}$ with $r$ as radial distance from the stellar disc
centre. The following parameter values are used unless otherwise
stated: an impact parameter of 0.5 stellar radii, a photon wavelength
of 6500~\AA{}, stellar radius equal to the solar value,
$\gamma_1=0.37$, and $\gamma_2=0.27$. The LDCs correspond to the solar
values computed using the theoretical tables of \citet{claret11} in
the \textit{Kepler} bandpass.

\subsection{Refraction model}\label{sec:refraction_model}
We present in the following an outline of the atmospheric refraction
transit light-curve model of \citet{hui02}, to which the reader is
referred for a rigorous derivation and comprehensive description of
the model. The model assumes isothermal atmosphere in hydrostatic
equilibrium and uniform vertical atmospheric composition. We note that
a factor of two is missing in Eq.~(9) of \citet{hui02}, see our
Eq.~\eqref{eq:u2phi}. We emphasise that it is the absolute magnitude
of the magnification ($A$) that enters Eq.~\eqref{eq:FF}. Negative
magnifications signifies that the image has been spatially reversed,
but only the photometric amplitude is of interest for refraction light
curves. \citet{hui02} also treat the effects caused by oblateness of
planets, but in this work planets are assumed to be spherical. In this
paper we use the notation shown in Table~\ref{tab:notation}, which is
the same as the notation used by \citet{hui02}.
\begin{table}
  \centering
  \caption{\label{tab:notation} List of symbols~\citep{hui02}.}
  \begin{tabular}{l l}\hline\hline
    Symbol & Meaning \\\hline
    $k_{\rm B}                                $ & Boltzmann constant                  \\
    $t                                        $ & time                                \\
    $T                                        $ & temperature                         \\
    $g                                        $ & surface gravity                     \\
    $\mu                                      $ & mean molecular weight               \\
    $m_{\rm H}                                $ & hydrogen atom mass                  \\
    $H                                        $ & atmospheric scale height            \\
    $\alpha                                   $ & refractive coefficient              \\
    $F                                        $ & normalised flux                     \\
    $W                                        $ & occultation kernel                  \\
    $D_\mathrm{OL}                            $ & observer--lens (observer--planet) distance  \\
    $D_\mathrm{LS}                            $ & lens--source (planet--star) distance  \\
    $\tau                                     $ & optical depth                       \\
    $R_0                                      $ & planet radius  \\
    $\rho_0                                   $ & atmospheric mass density at $R_0$   \\
    $\thetaS \equiv (\theta_\mathrm S^1, \theta_\mathrm S^2)$ & source position in the sky plane    \\
    $\thetaI \equiv (\theta_\mathrm I^1, \theta_\mathrm I^2)$ & image position in the sky plane     \\
    $\starangle                               $ & star-centre position in the sky plane              \\\hline
  \end{tabular}
\end{table}
Let $I(\thetaS - \starangle(t))$ be the surface brightness of the star
as a function of source position and star-centre position and
$W(\thetaI)$ be the occultation kernel of the planet disc, which is 0
when the observed image position is within the planet radius and 1
otherwise. \citet{hui02} verified that the step model was a good
approximation to reality. Then, the equations describing atmospheric
refraction become
\begin{equation}
  \label{eq:FF}
  F(t) = \int d^2\theta_\mathrm{S} \sum_\text{images} |A|\, I(\thetaS - \starangle(t))\, W(\thetaI)
\end{equation}
\begin{equation}
  \label{eq:AA}
  A = \left[ 1 + 2 \psi + \psi^2
+ u^2 \tilde\psi(1 + \psi)\right]^{-1}
\end{equation}
\begin{equation}
  \label{eq:phi}
  \psi(u) = - B \sqrt{\frac{\pi H}{2 u D_{\rm OL}}} {\,\rm exp}\left[-\frac{u
  D_{\rm OL} -R_0}{H}\right]
\end{equation}
\begin{equation}
  \label{eq:u2phi}
  u^2\tilde \psi(u) = B \sqrt{\frac{\pi u D_{\rm OL}}{2H}} 
  \left[1 + \frac{H}{2u D_{\rm OL}}\right] {\,\rm exp}\left[-\frac{u
      D_{\rm OL} -R_0}{H}\right]
\end{equation}
\begin{equation}
  \label{eq:uu}
  u^2 \equiv ({\theta_\mathrm I^1})^2 + ({\theta_\mathrm I^2})^2 = |\thetaI|^2;
\end{equation}
with parameters given by
\begin{equation}
  \label{eq:HH}
  H = \frac{k_{\rm B} T}{g \mu m_{\rm H}},
\end{equation}
\begin{equation}
  \label{eq:BB}
  B \equiv 2 \alpha \,\frac{\rho_0}{H} \frac{D_{\rm LS} D_{\rm OL}}{{D_{\rm OL} + D_{\rm LS}}}\approx 2\alpha\,\frac{\rho_0}{H}\,D_{\rm LS},
\end{equation}
and
\begin{equation}
  \label{eq:CC}
  C\equiv R_0/H,
\end{equation}
where the approximation $D_{\rm OL} \gg D_{\rm LS}$ in
Eq.~\eqref{eq:BB} holds for exoplanet transits. The transit light
curve is given by Eq.~\eqref{eq:FF} and is an integral over the
stellar disc, which is treated as a flat surface. The quantities
denoted by $\psi$ and $u^2\tilde\psi$ can be thought of as lensing
potentials. The parameter $B$ is essentially just a deflection angle
that has been scaled by a ratio of distances and $C$ is the ratio of
binding energy to thermal energy.

Much of the physics is captured by the important parameters $B$ and
$C$. Of
particular importance is the dividing line between strong and weak
lensing
\begin{equation}
  \label{eq:caustic}
  1-\sqrt{\frac{\pi}{2C}}B < 0,
\end{equation}
where the system is strongly lensing if the condition is
fulfilled~\citep{hui02}. Strong lensing occurs when caustics are
present. Caustics are source positions for which the magnification
diverges~\citep{hui02}. A more intuitive and equivalent condition for
strong lensing is when photons can be refracted into view by the far
side of the planet.  The far side is where the image is closer to the
planet centre than its source position. A consequence of strong
lensing is that the flux can increase above the baseline before and
after transit.  The baseline refers to the out-of-transit flux. The
increase of flux just outside of transit will henceforth be referred
to as `refraction shoulders'. Importantly, the information contained
in the $B$ and $C$ parameters does not capture all the properties that
could be connected to the photometric signal strength. Importantly,
refraction can obscure atmospheric layers only if the planet is
strongly lensing, which is crucial for transmission spectroscopy (see
Sect.~\ref{sec:transmission}).

Other models for refraction have been published based on a ray-tracing
approach~\citep{garcia_munoz12, betremieux13, misra14, betremieux15},
which implies that the equivalent to the analytic expression for the
magnification in Eq.~\eqref{eq:AA} needs to be determined
numerically. This approach is computationally more demanding than the
model of \citet{hui02}, which only requires Eq.~\eqref{eq:FF} to be
computed numerically. This is because the assumption of an atmosphere
that is isothermal in hydrostatic equilibrium and with uniform
vertical composition allows the use of an analytic expression in
Eq.~\eqref{eq:AA}. We note that \citet{misra14b} reported that the
assumption of an isothermal atmosphere results in no significant
difference after testing realistic tropospheric lapse rates and
stratospheric temperature inversions using their ray-tracing
framework. Alternately, \citet{sidis10} developed a model that
provides analytic approximations for the observed flux near and away
from occultation, but at the cost of ignoring limb darkening.

\subsection{Test planets}\label{sec:test_planets}
We construct a suite of test planets for the analysis of refraction
transit light curves. Each planet is defined by a set of parameters,
$T$, $M_\mathrm{P}$, $R_0$, $D_{\rm LS}$, and atmospheric composition,
where $M_\mathrm{P}$ is the planet mass and the remaining variables
are defined following \citet{hui02} (Table~\ref{tab:notation}). The
atmospheric composition is related to $\alpha$ by Cauchy's equation
(Appendix~\ref{sec:cauchy_coefficients}). Let $P$ be the orbital
period, $G$ the gravitational constant, $M_\mathrm{S}$ the stellar
mass, $\sigma_\mathrm{R}$ the cross section for Rayleigh scattering,
and assume the planetary orbits to be circular. The remaining
parameters needed to compute refraction light curves are then given
by~\citep{hui02}
\begin{equation}
  \label{eq:PP}
  P = 2\pi \sqrt{\frac{D_\mathrm{LS}^{3}}{GM_\mathrm{S}}},
\end{equation}
\begin{equation}
  \label{eq:gg}
  g = \frac{GM_\mathrm{P}}{R_0^2},
\end{equation}
\begin{equation}
  \label{eq:sig}
  \sigma_\mathrm{R} = 10^{-27}\text{~cm}^2\,\left(\frac{5000\text{~\AA{}}}{\lambda}\right)^4,
\end{equation}
\begin{equation}
  \label{eq:rho0}
  \rho_0 = \frac{\mu m_\mathrm{H}}{\sigma_\mathrm{R}\sqrt{2\pi R_0 H}},
\end{equation}
along with Eqs.~\eqref{eq:FF}--\eqref{eq:BB}.
Equation~\eqref{eq:rho0} assumes that the effective surface is set by
Rayleigh scattering. This is a reasonable assumption at a wavelength
of 6500~\AA{} in atmospheres devoid of clouds and hazes, which is what
is assumed here unless otherwise stated. Importantly, the model is
insensitive to the opacity source. The model is constructed such that
the optical depth above $R_0$ is zero and infinite below. So, it is
perfectly legitimate to ignore the Rayleigh scattering modelled by
Eq.~\eqref{eq:rho0} and to set $\rho_0$ and $R_0$ manually. For
example, if there are reasons to believe that a cloud deck obscures
all paths below $R_0$, then $R_0$ can be set to the radius that
corresponds to the altitude at the top of the cloud layer, in which
case $\rho_0$ would be the atmospheric mass density at $R_0$.

\begin{table*}
  \centering
  \caption{\label{tab:planet_params} Physical, atmospheric, and orbital parameters for the test planets.}
  \begin{tabular}{l c c c c c c c c c c}\hline\hline
    Planet      & Period           & Distance            & Mass             & Radius             & Temp.    & Atm. composition                         & $\alpha$          & $H$              & $B$  & $C$ \\
                    &                  &  (R$_\sun$)      &                  &                    & (K)     &                                    & (cm$^3$~g$^{-1}$) &     (km)         &      &     \\\hline
    Earth       & \hphantom{0}1~yr & \hphantom{0}215  & 1.0~M$_\oplus$     & 1.0~R$_\oplus$     & 255       & $0.78\mathrm N_2+0.22\mathrm{O}_2$ &                   0.23     & \hphantom{11}7.4 &            23048 & \hphantom{1}859 \\
    Super-Earth & 80~d\hphantom{r} & \hphantom{00}78  & 3.9~M$_\oplus$     & 1.5~R$_\oplus$     & 423       & $0.78\mathrm N_2+0.22\mathrm{O}_2$ &                   0.23     & \hphantom{11}7.1 & \hphantom{1}7302 &            1346 \\
    Jupiter     &            12~yr &            1118  & 1.0~M$_\mathrm{J}$ & 1.0~R$_\mathrm{J}$ & 150       & $0.86\mathrm H_2+0.14\mathrm{He}$  &                   1.22     & \hphantom{1}21.7 & \hphantom{1}3059 &            3276 \\
    Jovian      & \hphantom{0}1~yr & \hphantom{0}215  & 1.0~M$_\mathrm{J}$ & 1.0~R$_\mathrm{J}$ & 342       & $0.86\mathrm H_2+0.14\mathrm{He}$  &                   1.22     & \hphantom{1}49.4 & \hphantom{11}171 &            1437 \\
    Jovian      & 20~d\hphantom{r} & \hphantom{00}31  & 1.0~M$_\mathrm{J}$ & 1.0~R$_\mathrm{J}$ & 901       & $0.86\mathrm H_2+0.14\mathrm{He}$  &                   1.22     &            130.1 & \hphantom{0000}6 & \hphantom{1}546 \\
    Best-case   &            12~yr &            1118  & 1.0~M$_\mathrm{J}$ & 1.0~R$_\mathrm{J}$ & 600       &                     $\mathrm H_2$  &                   1.54     & \hphantom{1}98.5 & \hphantom{10}351 & \hphantom{1}721 \\\hline
  \end{tabular}
  \tablefoot{ Atmospheric composition fractions are given by
    number. The $0.86\mathrm H_2+0.14\mathrm{He}$ mixture for the
    Jovian planets is equivalent to 76\,\% H$_2$ and 24\,\% He by
    mass, which is the atmospheric composition of Jupiter.}
\end{table*}
\begin{figure}
  \centering
  \includegraphics[width=\hsize]{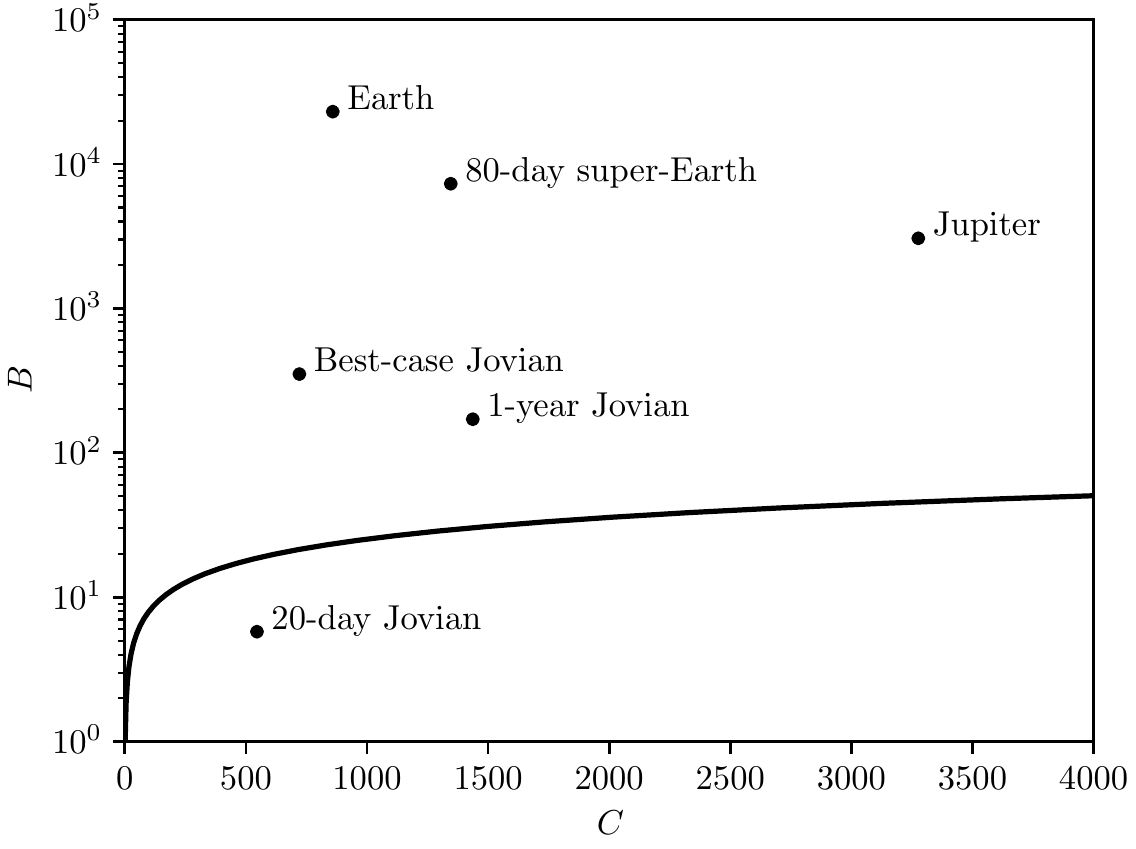}
  \caption{Positions of the test planets presented in
    Table~\ref{tab:planet_params}. The black line is the limit between
    weak and strong lensing; the region above the line is strongly
    lensing.\label{fig:bcp}}
\end{figure}
We provide the properties of all test planets in
Table~\ref{tab:planet_params}, while their distribution in the
$BC$-plane is shown in Fig.~\ref{fig:bcp}. Earth and Jupiter are used
as references for the other fiducial planets. The temperatures for
Earth and Jupiter are chosen to be representative of the majority of
the atmosphere above the effective Rayleigh surface. A temperature of
255~K is deemed representative of the atmosphere of the Earth and the
temperature of 150~K for Jupiter is based on the temperature profile
measured by the \textit{Galileo} atmospheric
probe~\citep[][Fig.~28]{seiff98}. The temperature of the 80-day
super-Earth is assumed to scale as $T\propto D_\mathrm{LS}^{-0.5}$
with respect to the Earth and the temperatures of the 20-day Jovian
and 1-year Jovian with respect to Jupiter. The bulk density for the
fiducial super-Earth scenario is chosen based on the empirical
relation of \citet[][Eq.~3]{weiss14}, which is fitted to planets with
radii in the range 1.5--4~R$_\oplus$. This relation results in a mass
of 3.9~M$_\oplus$ for a radius of 1.5~R$_\oplus$, corresponding to a
mean density of 6.4~g~cm$^{-3}$. It is also assumed that the mass
distributions within all planets are homogeneous. The average density
of the Jovian planets is assumed to be equal to that of Jupiter.

The purpose of the best-case planet is to put a limit on the maximum
possible refraction strength. We fix the orbital distance to the
orbital distance of Jupiter because it sets the geometrical deflection
angle and is of little importance as long as the angle is small
(Sect.~\ref{sec:refraction_shoulders}). The radius is set to the
Jupiter radius because an arbitrarily large signal strength can be
achieved by increasing the radius with an appropriate rescaling of the
other parameters. This leaves the atmospheric composition, which is
set to pure H$_2$; planet mass, which is set to 1~M$_\mathrm{J}$; and
the temperature, which is set to 600~K. It is shown in
Sect.~\ref{sec:refraction_shoulders} that the atmospheric composition,
planet mass, and temperature are degenerate. For example, a doubling
in mass and temperature results in an identical light curve to that of
the original parameters. The purpose of the best-case planet is that
any planet at an orbital distance and with a radius equivalent to
those of Jupiter yields a weaker refraction signal. In this sense, the
best-case planet scenario can be considered as an upper limit. We note
that the Earth happens to show a stronger refraction signal for an
Earth-sized planet on a 1-year orbit and can serve as a best-case
terrestrial planet.

\section{Model predictions}\label{sec:mod_results}
\subsection{Refraction light curves}
\begin{figure}
  \centering
  \includegraphics[width=\hsize]{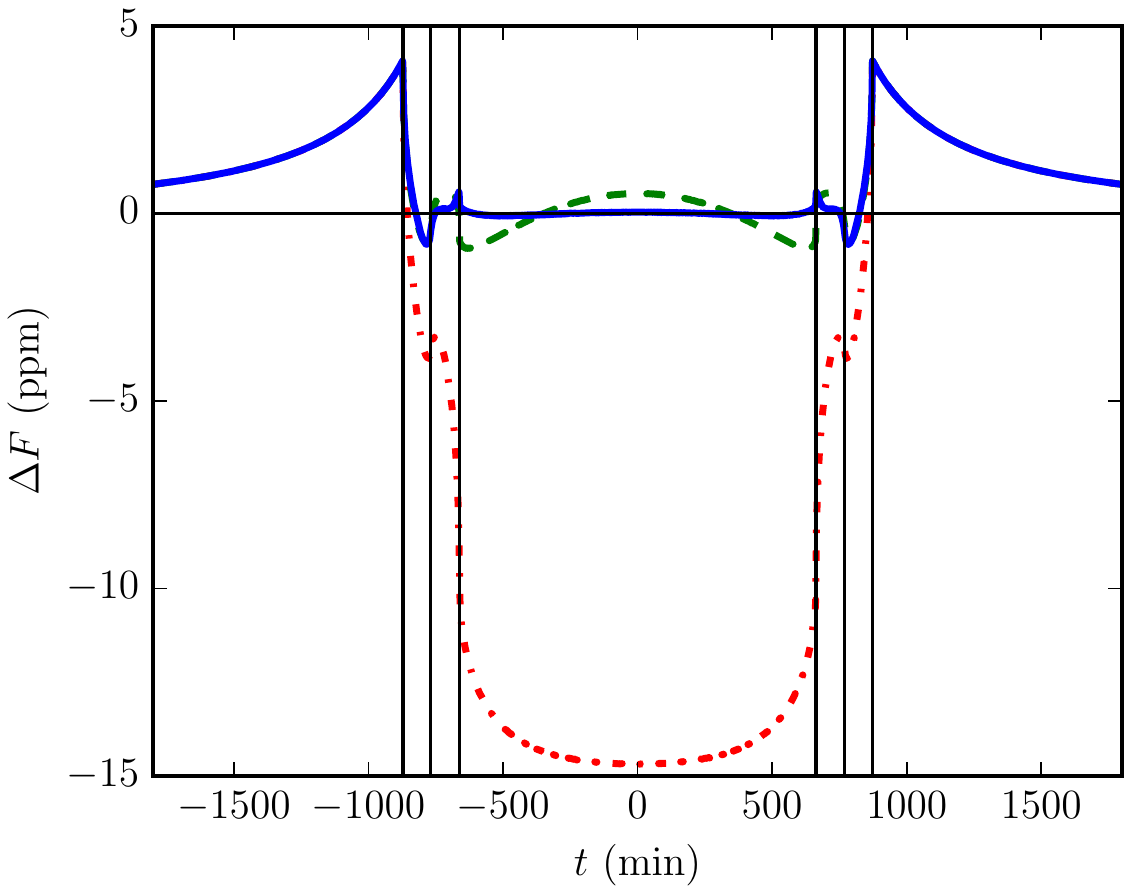}
  \caption{Residuals for a light-curve model including refraction
    fitted with a plain transit model without refraction for
    Jupiter. The dash-dotted red line is for a plain model with
    identical parameters as for the refraction light curve. The dashed
    green line is for a plain model with fixed, correct limb-darkening
    coefficients (LDCs). The solid blue line is a plain model fitted
    with 10\,\% freedom in the LDCs. The vertical black lines mark
    when the planet limb and centre cross the star limb. The reference
    time is set to mid-transit.\label{fig:jupiter_fit_res}}
\end{figure}

The primary question is what differences are caused by atmospheric
refraction in transit light curves. This can be studied by generating
a light curve using the refraction model of \citet{hui02} and then
fitting a model without refraction to it. Residuals ($\Delta F$) can
then be computed by taking the refraction light curve and subtracting
the fitted model from it.  Mathematically,
$\Delta F = F_\mathrm{R}-F_\mathrm{M}$, where $F_\mathrm{R}$ is the
normalised predicted model flux and $F_\mathrm{M}$ is the normalised
flux of a fitted model. Figure~\ref{fig:jupiter_fit_res} shows
residuals for a refraction light curve fitted with plain transit
light-curve models without refraction for Jupiter. The fitted models
are a plain transit light curve with the parameters fixed to those
used for generating the refraction light curve (dash-dotted red); with
free inclination, orbital distance, and planet radius but fixed LDCs
(dashed green); and with free parameters and 10\,\% freedom in the
LDCs (solid blue). The effects of refraction are clearly seen as
shoulders before and after transit and a flux decrease appears during
transit. The shoulders only exist for planets in the strong lensing
regime. The near side of a strongly lensing planet can occult the star
before light starts being lensed into view by the far side of the
planet, in which case the shoulders are suppressed. This is only
relevant for systems close to the limit between strong and weak
lensing. Most systems in the strong lensing regime will exhibit
shoulders. The flux decrease of $\sim$15~parts per million (ppm,
dash-dotted red) at mid-transit for Jupiter is relatively large. For
comparison, the corresponding flux decrease for the 1-year Jovian case
is $\lesssim 1$~ppm. One of the main points of
Fig.~\ref{fig:jupiter_fit_res} is to show that many of the in-transit
refraction features are eliminated when fitting a model with free
impact parameter, planet radius, and star--planet distance (dashed
green), especially when also allowing for some freedom in the LDCs
(solid blue).

\subsection{Refraction shoulders}\label{sec:refraction_shoulders}
\begin{figure}
  \centering
  \includegraphics[width=\hsize]{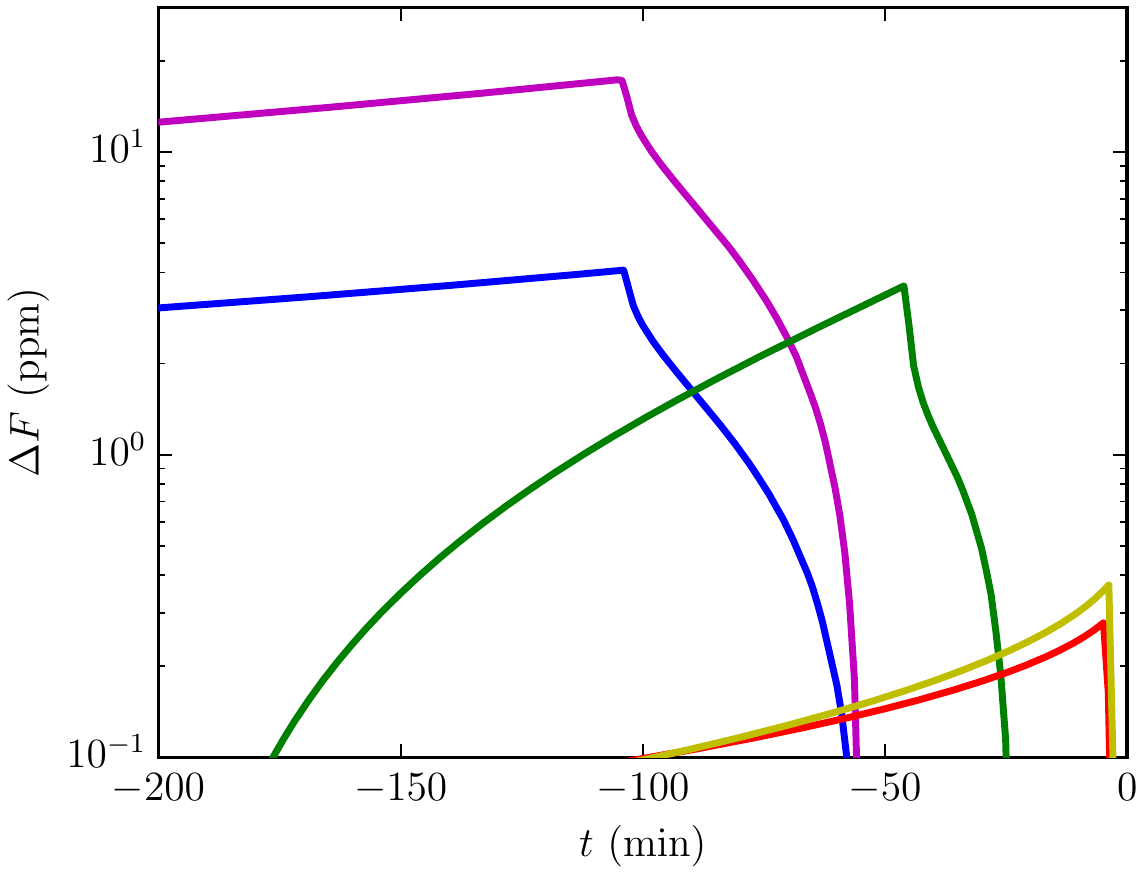}
  \caption{Refraction shoulders for the five strongly lensing test
    planets. The $y$-axis is logarithmic for visual clarity. Blue is
    the reference Jupiter and is identical to the blue line in
    Fig.~\ref{fig:jupiter_fit_res}. Red is the Earth, yellow is the
    80-day super-Earth, green is the 80-day Jovian, and magenta is the
    best-case Jovian planet. The reference time is set to when the
    planet centre crosses the star limb, which is equivalent to the
    second vertical line in
    Fig.~\ref{fig:jupiter_fit_res}.\label{fig:sample_wings}}
\end{figure}
Refraction shoulders are expected to be the most easily detectable
features and depend on several atmospheric parameters.
Figure~\ref{fig:sample_wings} shows refraction shoulders for the suite
of test planets provided in Table~\ref{tab:planet_params}, except for
the 20-day Jovian, which does not exhibit shoulders as it is weakly
lensing. The peak residual amplitudes and half-times for the planets
are provided in Table~\ref{tab:ref_par}. A half-time is defined as the
time interval measured from when the residual amplitude is half of the
peak amplitude to the time when the residual amplitude is at its peak.
\begin{table}
  \centering
  \caption{Refraction shoulder properties.\label{tab:ref_par}}
  \begin{tabular}{l l c c}\hline\hline
    Host  & Planet         &                Peak amp.      & Half-time \\
          &                &                (ppm)          & (min) \\\hline
    Sun   & Earth          &               \hphantom{0}0.3 &  \hphantom{0}49 \\
    Sun   & Super-Earth    &               \hphantom{0}0.4 &  \hphantom{0}33 \\
    Sun   & Jupiter        &               \hphantom{0}4.1 &             274 \\
    Sun   & 1-year Jovian  &               \hphantom{0}3.6 &  \hphantom{0}38 \\
    Sun   & Best-case      &                          17.3 &             221 \\
    M4V   & Earth          &               \hphantom{0}2.9 &  \hphantom{0}12 \\
    M4V   & Super-Earth    &               \hphantom{0}2.6 &  \hphantom{00}6 \\
    M4V   & Jupiter        &                          18.3 &  \hphantom{0}43 \\
    TRAPPIST-1 & b         &      \hphantom{0}2.0 &  \hphantom{00}1 \\ 
    TRAPPIST-1 & c         &      \hphantom{0}3.3 &  \hphantom{00}2 \\ 
    TRAPPIST-1 & d         &      \hphantom{0}7.4 &  \hphantom{00}2 \\ 
    TRAPPIST-1 & e         &                 19.6 &  \hphantom{00}3 \\ 
    TRAPPIST-1 & f         &                 23.8 &  \hphantom{00}6 \\ 
    TRAPPIST-1 & g         &                 18.7 &  \hphantom{00}9 \\ 
    TRAPPIST-1 & h         &                 40.2 &  \hphantom{00}8 \\\hline
  \end{tabular}
\end{table}
Systems with higher peak amplitudes are not
necessarily easier to detect. For instance, when comparing the 80-day
super-Earth with the Earth, the super-Earth's refraction shoulders have a higher peak
amplitude, but fade over shorter timescales than the shoulders of the Earth.

We explore the parameter space by studying the strength of refraction
as a function of the following parameters: observing wavelength,
temperature, mass, radius, orbital distance, and molecular
weight. This is done by scaling each parameter with respect to the
reference value that is provided in Table~\ref{tab:planet_params}. The
molecular weight is scaled by simply assuming that the mass of the
atoms that constitute the atmosphere is changing. This means that the
refractive properties, which are unique for each molecule, are assumed
to be constant. Each parameter is assumed to be independent of the
other parameters that are used to define a test planet, but propagates
into the quantities that are calculated using the scaled parameter,
following the description in Sect.~\ref{sec:test_planets}. For
example, the mass is assumed to be constant when the radius is
doubled, whereas a parameter such as the surface gravity would be
reduced by a factor of 4 following Eq.~\eqref{eq:gg}. We thus define
the refraction signal strength $S_{\!\Delta t}$ as the mean $\Delta F$
over the $\Delta t$ minutes just before first contact. This serves as
a measure of the shoulders' detectability.

\begin{figure*}
  \centering
  \includegraphics[width=0.48\hsize]{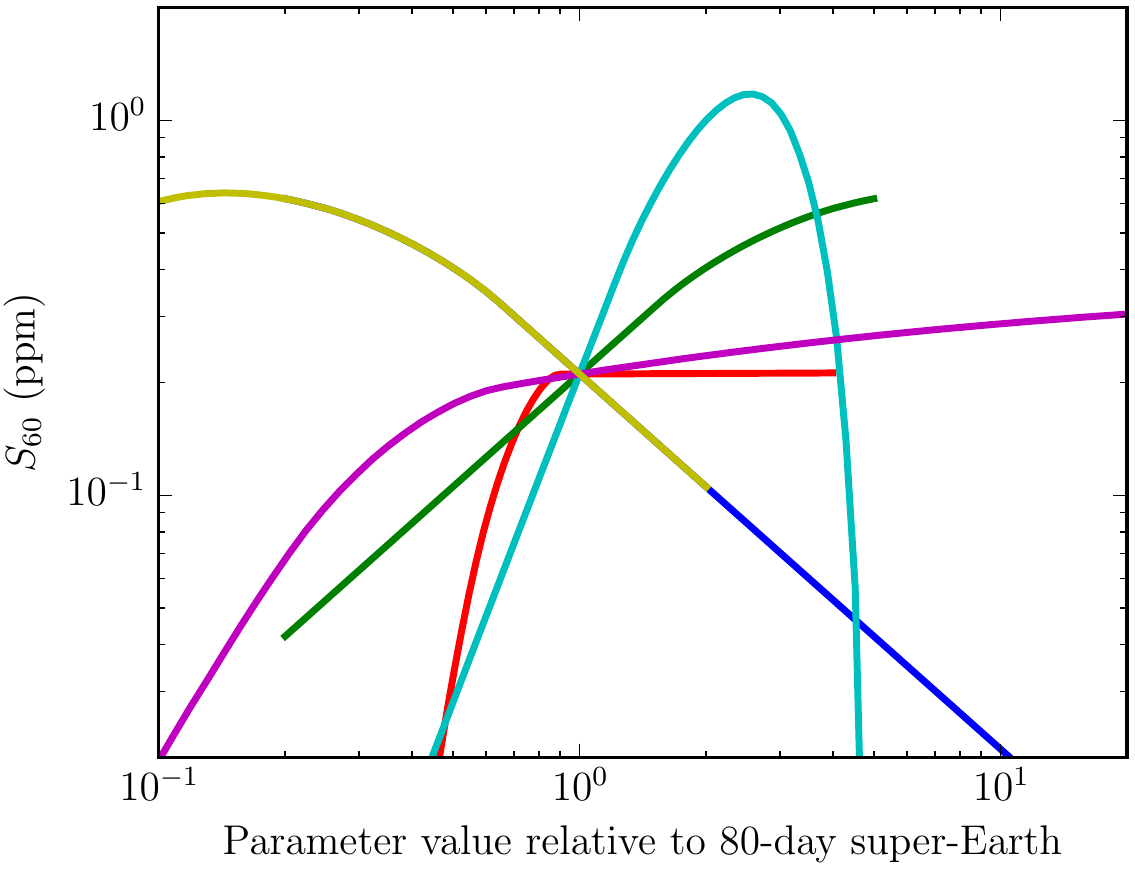}
  \hfill{}
  \includegraphics[width=0.48\hsize]{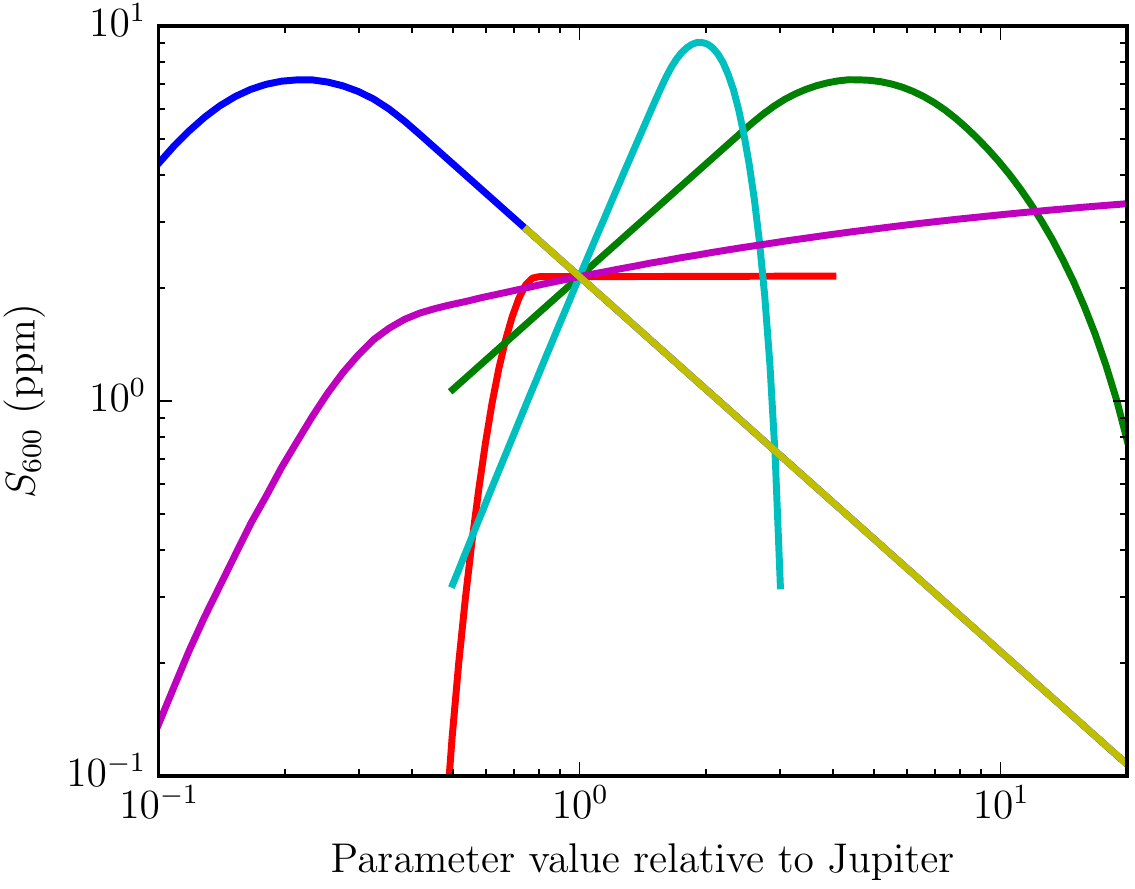}
  \caption{Refraction signal strengths as functions of six different
    refraction parameters. The left panel is the 80-day super-Earth
    and the right panel is Jupiter. The $x$-axis is the ratio of the
    changing parameter to its original value. Red is wavelength, green
    is temperature, blue is planet mass, cyan is radius, magenta is
    orbital distance, and yellow is molecular weight (which overlaps
    with planet mass). The signal is measured over 60~min for the
    80-day super-Earth and 600~min for Jupiter because of the
    different characteristic timescales.\label{fig:mod_run}}
\end{figure*}
Refraction signal amplitudes for various ranges of parameters are
shown in Fig.~\ref{fig:mod_run}. The very steep refraction strength
drop-off at shorter wavelengths is primarily explained by the
fourth-order wavelength dependence of Rayleigh scattering (see
Eq.~\eqref{eq:sig}). The radius $R_0$ is set to the radius at which
the optical depth due to Rayleigh scattering is unity. A quickly
increasing scattering cross section at shorter wavelengths implies
that the effective radius is shifted to higher altitudes where the
density is lower. A secondary effect of decreasing wavelength is that
the refractive coefficient increases, but this effect is orders of
magnitude weaker than that of the increased Rayleigh cross
section. The wavelength dependence flattens for long wavelengths
because it only determines the capability of the atmosphere to deflect
light.  This enters the model through the $B$ parameter, which is a
deflection angle scaled by distance. Further increased deflection
strength beyond the point where the planet is capable of deflecting
light from its host star into the line of sight does not result in a
significantly stronger signal because the limiting factor becomes the
projected area of the atmosphere. This means that $B$ is effectively
saturated.

Figure~\ref{fig:mod_run} also shows that the refraction signal
strength monotonically increases with increasing orbital distance. The
signal strength increases more steeply for shorter distances, then
reaches a break and keeps increasing more slowly. The steep region is
where the refraction strength is limited by the capability of the
atmosphere to deflect the light at a large enough angle. The break is
the distance at which the $B$ parameter effectively saturates,
analogously to the break that is seen in the wavelength
dependence. The reason for a continued, slower refraction strength
increase for large distances is that the transit timescale lengthens
as a consequence of the Keplerian orbits of the planets.

The refraction signal strength has a non-monotonous dependence on some
of the parameters. The signal shows the same behaviour with respect to
the molecular weight, mass, and inverse temperature because these
three parameters enter the equations through $B$ and $C$ identically.
The parameters decrease the scale height and increases the density
$\rho_0$, both of which result in a larger density gradient of the
atmosphere, which increases the atmosphere's capability of deflecting
light. A competing effect is that a decreased scale height also gives
a smaller projected atmospheric area, which results in a lower
refraction signal. The balance between these two effects yields a
parameter value for which the refraction strength is maximised. The
density has a stronger dependence on molecular weight than the planet
mass and inverse temperature, but this effect is cancelled by a
corresponding decrease in the refractive coefficient $\alpha$.

The strongest dependence of refraction amplitude is on the
planet-to-star radius ratio. The refraction signal scales with planet
radius because of the increased projected atmospheric area. Additional
effects are that the scale height increases and atmospheric density
decreases, which at a given radius renders the atmosphere unable to
deflect light at large enough angles to be strongly lensing, resulting
in a sharp cut-off. It is also possible that the radius is large
enough for the near side of the planet to start occulting the star
before the far side of the planet starts refracting light into
view. In this case, the strength of the refraction signal also drops
sharply.

\subsection{Red dwarf host star}\label{sec:red_dwarf}
\begin{figure}
  \centering
  \includegraphics[width=\hsize]{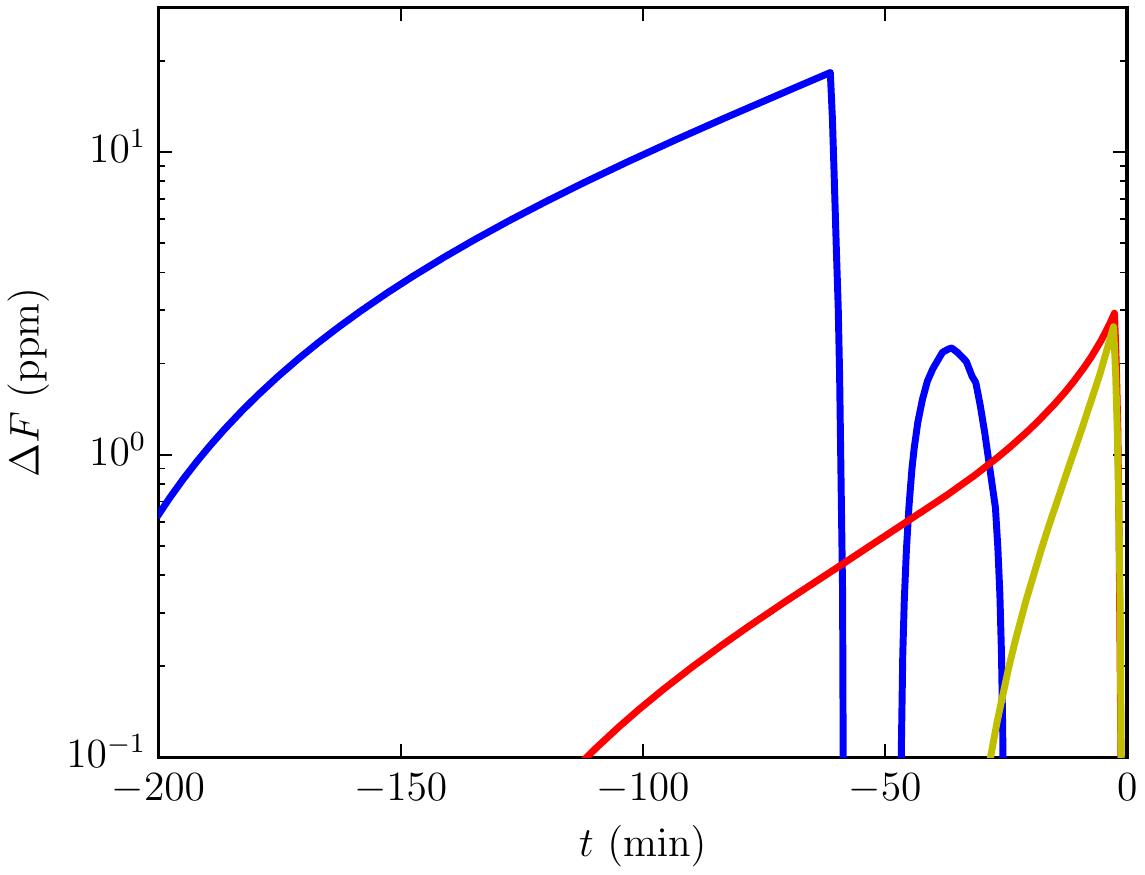}
  \caption{Same as Fig.~\ref{fig:sample_wings}, but with an M4V red
    dwarf host star. The 1-year Jovian and best-case planet are
    excluded;  they have no shoulders because they fall outside of
    the strong lensing regime.\label{fig:red_dwarf}}
\end{figure}
Refraction signal strength is dependent on host star properties, which
is investigated by computing the refraction wings for the same test
suite of planets but with a red dwarf host. The host star used is a
fiducial M4V dwarf star with a radius of 0.26, mass of 0.2, and
luminosity of 0.0055 relative to the
Sun~\citep{reid05,kaltenegger09}. The LDCs are set to $\gamma_1=0.2$
and $\gamma_2=0.4$ corresponding to the redder wavelengths of
\citet{berta-thompson15} because the flux of a red dwarf is assumed to
be weighted towards longer wavelengths in the \textit{Kepler}
bandpass. The test planets are moved to an orbit at which they receive
the same incident flux as for a Sun-like host. Only Earth, the 80-day
super-Earth, and Jupiter out of the five test planets are in the
strongly lensing regime with the M4V dwarf host. The refraction
shoulders are shown in Fig.~\ref{fig:red_dwarf}. The two most striking
differences are that the amplitudes increase by a factor of ${\sim} 5$
and the relevant timescales are much shorter, which is clearly seen in
Table~\ref{tab:ref_par}.

We can then compare the detectability of refraction features for
Jupiter around the Sun and a Jupiter analogue around an M4V dwarf
star. Let $\xi$ be the timescale, $L$ host star luminosity, and assume
that the achievable photometric precision ($\sigma_\mathrm P$) scales
as the square root of photon count. The precision is the
  relative uncertainty from photon counting statistics that can be
  achieved within a given time interval. For this reason a factor of
  $P^{-1/2}$ is introduced because it is assumed that observations are
  performed whenever a transit occurs. In this case the precision
  follows
\begin{equation}
  \label{eq:precision}
  \sigma_\mathrm P \propto \sqrt{\frac{\xi L}{P}}.
\end{equation}
The period around the dwarf star is six months compared to twelve
years around the Sun and the refraction timescale is shorter
  by a factor of $274/43\approx 6.4$ (from Table~\ref{tab:ref_par})
around the dwarf star. The ratio of precision around the Sun to
precision around an M4V dwarf after twelve years of observations is
then
\begin{equation}
  \label{eq:precision_ratio}
  \sqrt{\frac{274}{43}\times\frac{1}{0.0055}\times\frac{0.5}{12}} \approx 7.
\end{equation}
The signal strength around the Sun is expected to be weaker by roughly
the same amount, see Table~\ref{tab:ref_par}, so the
predicted signal-to-noise ratio (S/N) is similar for the Sun and for an
M4V dwarf host. Importantly, the M4V S/N would of course decrease if not
  all transits are observed within the given time interval.

\begin{figure}
  \centering
  \includegraphics[width=\hsize]{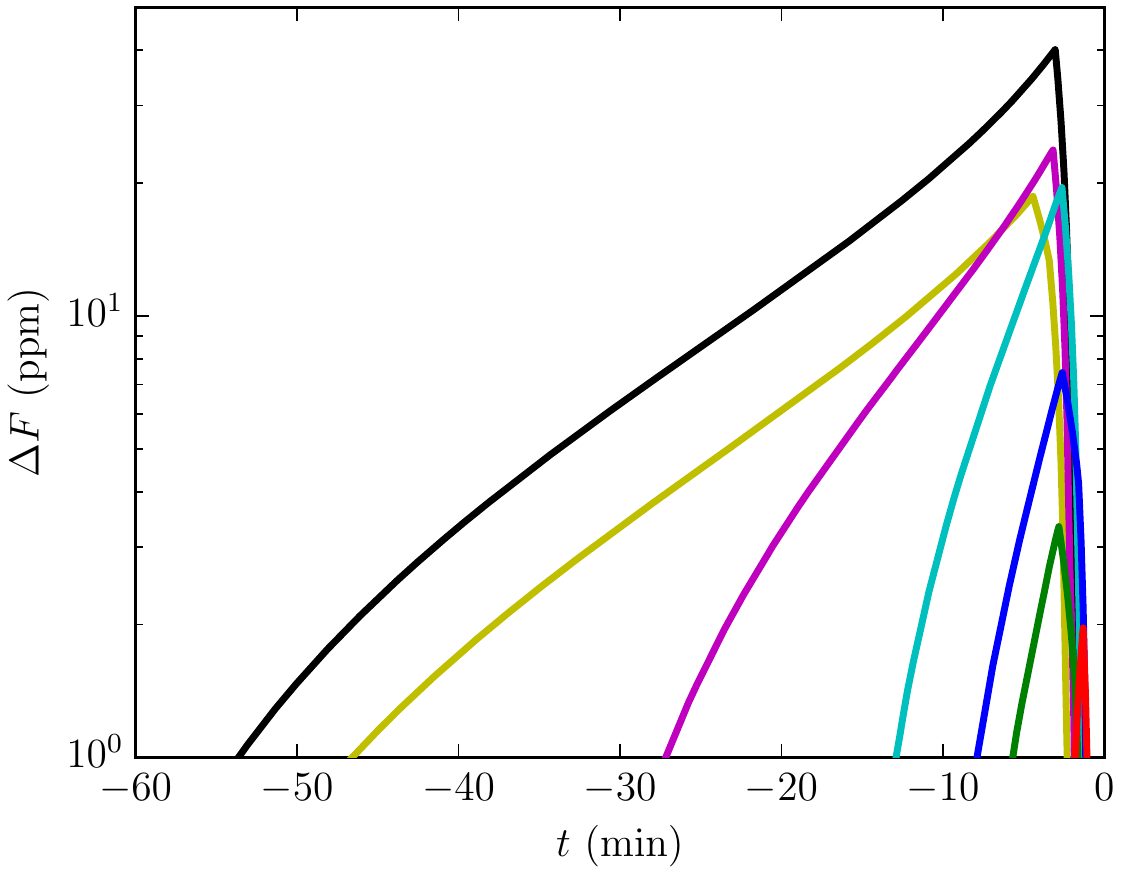}
  \caption{Same as Fig.~\ref{fig:sample_wings}, but for the TRAPPIST-1
    system at a wavelength of 4.5~$\mu{}$m corresponding to the
    infrared \textit{Spitzer} photometry of \citet{gillon17}. The
    planets are TRAPPIST-1b (red), c (green), d (blue), e
    (cyan), f (magenta), g (yellow), and h
    (black).\label{fig:trappist}}
\end{figure}
We now focus on the more extreme case of the M8V dwarf TRAPPIST-1, which hosts
  seven Earth-size planets~\citep{gillon17}. We compute the refraction
  shoulders for all TRAPPIST-1 planets at a wavelength of
  4.5~$\mu{}$m, corresponding to the infrared \textit{Spitzer}
  photometry gathered for this object \citep{gillon17}. The nominal effect of limb darkening at
  4.5~$\mu{}$m is ignored. It is not reasonable to assume that
  Rayleigh scattering sets the effective surface at a wavelength of
  4.5~$\mu{}$m. Instead, the planet radius ($R_0$) is set to the
  radius at which the pressure is 1~atm, consequently the mass density
  $\rho_0$ is set to the density at $R_0$, instead of being computed
  using Eq.~\eqref{eq:rho0}. We show the refraction shoulders for all TRAPPIST-1 planets in Fig.~\ref{fig:trappist} and their properties in Table~\ref{tab:ref_par}. We use the parameters (except for planet masses) from
  \citet{gillon17}, an assumed atmospheric composition $0.78\mathrm N_2+0.22\mathrm{O}_2$, and planet masses from \citet{wang17b}. The seventh planet, TRAPPIST-1h,
  shows the strongest refraction signal with a peak excess amplitude
  of 40~ppm and a half-time of 11~min. It is not reasonable to expect
  any residual signal to be stronger than 100~ppm for different
  assumptions concerning the atmospheric composition or when
  considering the uncertainty of the parameters. This value was determined
  by exploring the parameter space around TRAPPIST-1h.

The S/N for planets around TRAPPIST-1 and the Sun can be
  compared using the precision defined by Eq~\eqref{eq:precision}.
The luminosity of TRAPPIST-1 is 0.000524~L$_\sun$ and the orbital
period of TRAPPIST-1h is 18.8~days~\citep{luger17}. The
distance with an equivalent incident flux around the Sun would
correspond to a period of 1677~days. The peak amplitude of the
  refraction shoulders of a TRAPPIST-1h analogue around the Sun is
  1~ppm with a half-time of 73~min. Analogously to 
  Eq.~\eqref{eq:precision_ratio}, the precision would be higher by a
  factor of \begin{equation}
  \label{eq:precision_ratio_trappist}
  \sqrt{\frac{73}{8}\times\frac{1}{0.000524}\times\frac{18.8}{1677}} \approx 14
\end{equation}
around the Sun than TRAPPIST-1, which results in a comparable S/N given
that the refraction signal is 1~ppm around the Sun
compared to 40~ppm around TRAPPIST-1.

\subsection{Weak lensing}\label{sec:weak_lensing}
\begin{figure}
  \centering
  \includegraphics[width=\hsize]{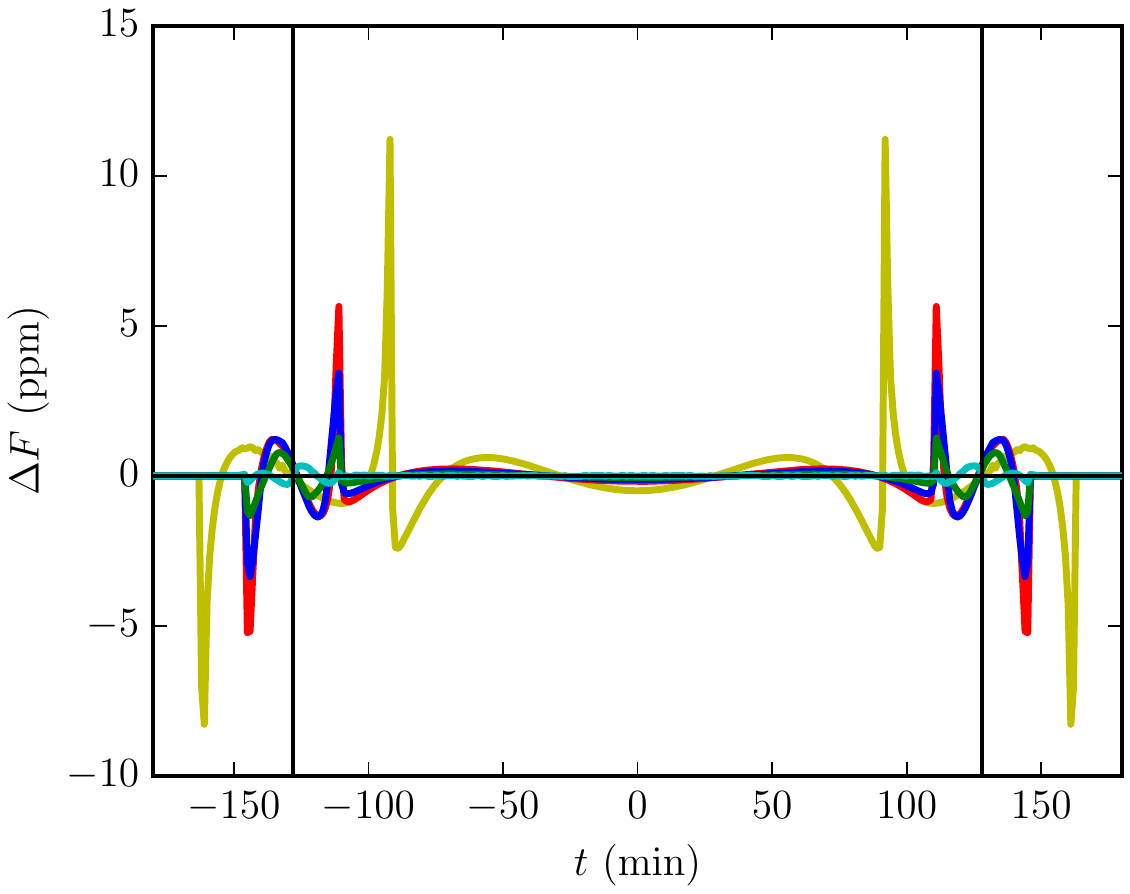}
  \caption{Residuals for a set of 20-day Jovians after fitting a plain
    light-curve model with 10\,\% freedom in the limb-darkening
    coefficients to a refraction light curve. Blue is the reference
    planet with parameters given in Table~\ref{tab:planet_params}. Red
    is double temperature, green is double mass, cyan is six times
    higher molecular weight, and yellow is double radius. The vertical black
    lines mark when the planet centre crosses the star limb. The
    reference time is set to
    mid-transit.\label{fig:20d_jovian_param_space}}
\end{figure}
Any in-transit, ingress, and egress signals are expected to be
vanishing for weakly lensing planets because almost all of these
refraction effects can be compensated for by fitting a plain transit
model, especially when allowing for some freedom in the LDCs. This can
be seen for Jupiter in Fig.~\ref{fig:jupiter_fit_res}, where the
residuals are smaller than 1~ppm between the shoulders when fitting a
model with 10\,\% freedom in the LDCs. The 20-day Jovian is a planet
that is expected to have relatively strong lensing features even
though it is weakly lensing, primarily because it is large and is
further away from its host star than the closest close-in
giants. Figure~\ref{fig:20d_jovian_param_space} shows the residuals
for the 20-day Jovian planet along with a set of planets with similar
parameters. The main point is that refraction features are not
significantly stronger and present on much shorter timescales compared
to strongly lensing planets.

\section{Observations}\label{sec:observations}
We then turn to observations where transit light curves from \textit{Kepler} are stacked and fitted in an
attempt to detect refraction features. \textit{Kepler}
  30-minute long-cadence simple aperture photometry (SAP)~\citep{jenkins10}
   (data release 25) are used for the analysis. The products in
  the SAP data used for our purpose are the timestamps, photometric fluxes,
  1-$\sigma$ statistical flux errors, and quality flags. Data for all
Kepler objects of interest (KOIs) from quarters 0 through 17 were
downloaded from NASA Exoplanet
Archive\footnote{\url{http://exoplanetarchive.ipac.caltech.edu}} (NEA)
on 23 January 2017. The SAP data have Argabrightening events
  mitigated~\citep{witteborn11} and cosmic rays removed. Background
  estimates are then subtracted and the photometric fluxes are
  computed using optimal apertures. The SAP is a less processed version
  than the pre-search data conditioning SAP (PDCSAP)~\citep{stumpe12, smith12},
   which has undergone additional artefact mitigation to
  account for instrument systematics. The choice of using SAP instead
  of PDCSAP is made because the artefacts primarily affect longer
  timescales and additional high-level data reduction increases the risk of
  either removing refraction signals or introducing weak systematics that could be
  confused with refraction-driven patterns in the data. Concerning false positives, \citet{fressin13} determined that 90\,\% of all
KOIs are real planets.

\subsection{Light curve reduction and fitting}\label{sec:reduction}
The \textit{Kepler} photometry data are reduced and fitted in order to
study any systematic residuals. All transits for each KOI are stacked
and only data corresponding to five transit durations centred on the
transit are kept for each transit. All data points during transits of
other KOIs in the same system or flagged as bad in the SAP are
removed. Bad means that the flag \texttt{SAP\_QUALITY} is different
from 0. This conservative rejection criterion is chosen because
systematic residuals were observed otherwise. Each transit is
  individually normalised by dividing the data with a quadratic
  polynomial that is fitted to the out-of-transit baselines, which are
  defined as the first and last 1.5 transit durations out of the five
  transit durations. This means that the actual transit and half a
  transit duration before and after the transit are excluded. The
  additional margin is introduced because of potential refraction
  shoulders during the time intervals just before and after
  transit. The transit
light-curve model of \citet{mandel02} is then fitted to the stack of
all transits for each KOI. Transit timing variations (TTVs) are modelled by introducing a sinusoidal shift in time. Light curve fitting is performed
  using the Levenberg--Marquardt gradient descent function
  \texttt{curve\_fit} in SciPy 0.13.3 with eight free parameters;
  star--planet distance, planet radius, inclination, two LDCs and three
  TTV parameters. The LDCs are given
10\,\% freedom with respect to the values in NEA, which are derived
using the method of \citet{claret11} using the ATLAS model and least-squares method with effective temperatures and surface gravities from
\citet{huber14}. The three TTV parameters are the amplitude,
  frequency, and initial phase of the sinusoidal shift in time. A
5-$\sigma$ clip is then applied to the data meaning that points with a
photometric residual of more than five sigma are removed.
A second normalisation using the same conservative baseline is then
performed with the additional requirement that the normalisation
function is fitted to at least five points on each side. A minimum of
14 data points between the baseline intervals are demanded to ensure
that the reference level is well determined and that the transit
contains a meaningful amount of information. Transits with too few
points are removed and the remaining stacked transits are refitted.

The accuracy of the photometric baseline for each transit is checked
using two tests~\citep{sheets14}. Firstly, a line is fitted to the
first interval and projected to the other side of transit. The transit
is rejected if the mean of the projection at the time of the other
baseline deviates by more than 0.001 (relative flux) from 0. It was
verified that the choice of 0.001 resulted in a reasonable balance
between number of rejected transits and light-curve quality. The
projection is then reversed by fitting to the second interval and
projecting to the first. It is possible that a transit light curve
passes one but not both parts of the test. Secondly, it is required
that the two lines from the projection test have slopes that are
consistent with 0 at 3-$\sigma$. Only transits with both slopes being
consistent with 3-$\sigma$ are kept for further analysis. Light curve
fits are then performed for each KOI to the stack of the transits that
passed both tests. Finally, the deviations of the residuals are tested
for systematics using the red-noise test presented
in~\citet{sheets14}. If the noise is solely composed of statistical
fluctuations, such as for photon counting, then
$\log(\sigma) \propto -0.5\log(N)$, where $\sigma$ is the sample
standard deviation and $N$ is the size of the bins. The stacked
residuals for all transits of a given KOI are binned with different
bin sizes, allowing for fitting of $\log(\sigma)$ against
$\log(N)$. Only KOIs with fitted noise slopes in the range $-0.8$ to
$-0.3$ are kept. This is an efficient way to filter KOIs with various
systematic errors on different timescales.

A total of 2394 out of the 4707 KOI light curves pass all tests. The
median reduced $\chi^2$ value for the 2394 fits is 1.24 and the
arithmetic mean is 1.47. The distribution of the number of degrees of
freedom (DOF) for the 2394 planets has a median of 8680 and arithmetic
mean of 13\,733. The primary condition that candidates failed was the
requirement of 14 data points between the intervals used for baseline
fitting. This is essentially equivalent to requiring transits to be
longer than three hours, which excludes 1743 KOIs. These are close-in
planets and are not expected to show any refraction signal because of
the geometry of the planetary system and the refraction timescale
compared to the 30-minute cadence of \textit{Kepler}. This scenario is
studied in Sect.~\ref{sec:weak_lensing}; examples are shown in
Fig.~\ref{fig:20d_jovian_param_space}.

\subsection{Selection of candidates}
A subset of 305 out of the 2394 fitted KOIs are selected based on the
following criteria. It is required that the planet radius is larger
than 0.025 stellar radii and orbital period longer than 40 days. This
is based on theoretical expectations of which planet population would
show the strongest refraction signals. The choice of having an
orbital-period requirement instead of star--planet distance cut-off is
because the period is more robust as it is measured directly in
contrast to the star--planet distance, which is determined by the
multi-dimensional light-curve fit. It is required that fitted planet
radii should be smaller than 0.5 of the host star radius and that
impact parameters be smaller than 1. These criteria mainly reject
poorly fitted KOIs. The subset of 305 large planets with long orbits
will henceforth serve as the primary sample in the analysis. The
median reduced $\chi^2$ value for the 305 selected planets is 1.22 and
the arithmetic mean is 1.46. The distribution of the DOF for the 305
planets has a median of 3358 and arithmetic mean of 3941.

\section{Stacked \textit{Kepler} data}\label{sec:obs_results}
\begin{figure}
  \centering
  \includegraphics[width=\hsize]{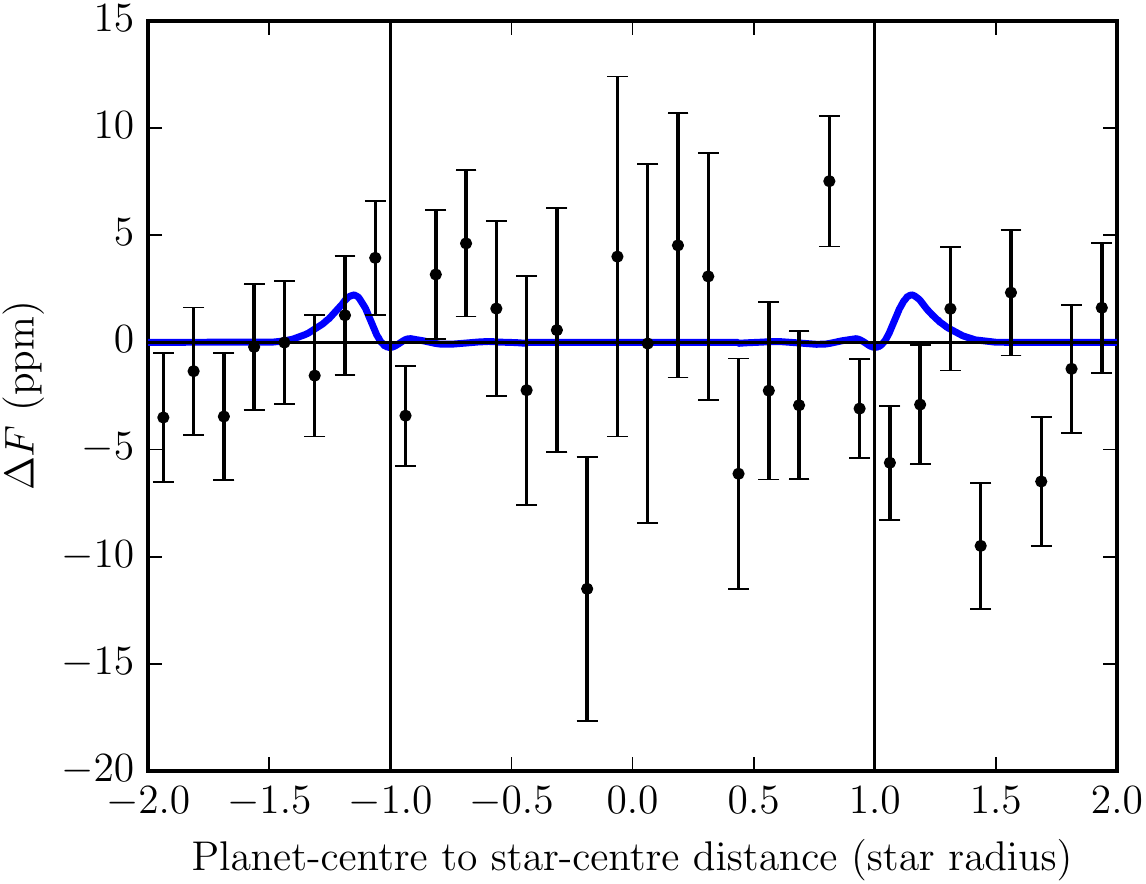}
  \caption{Binned residuals for the sample of 305 large planets on
    long orbits. The blue line is the theoretical prediction for a
    Jovian planet on a 1-year orbit. The precision of the stacked
    residuals is clearly not high enough to detect any refraction
    signal. Negative distances are before mid-transit and positive distances are after. The vertical black lines mark when the planet centre
    crosses the star limb.\label{fig:stack_wings}}
\end{figure}
Figure~\ref{fig:stack_wings} shows the stacked residuals for the
selected sample of 305 large planets with long orbits. A total of
101\,557 long-cadence data points are stacked, with an average of 3174
points per bin. The bins are 0.125 wide in units of host star
radii. The relatively long timescale is chosen to reveal refraction
shoulders. The residual signal is expected to be of the order of
1~ppm. The precision of the stacked residuals is in the range
2.3--8.4~ppm, implying that any refraction shoulders are not expected
to be detectable. The increased noise towards mid-transit occurs primarily
because many projected KOI transits do not cross the central parts of
the host star, resulting in fewer data points. The blue
line is the theoretical prediction for a Jovian planet on a 1-year
orbit.  It is not a best-case scenario, but a signal in the averaged
residuals can most likely not be expected to be stronger than that of
a 1-year Jovian. This is determined by exploring the parameter
  space and studying the residuals for many planets in the sample.

\begin{figure}
  \centering
  \includegraphics[width=\hsize]{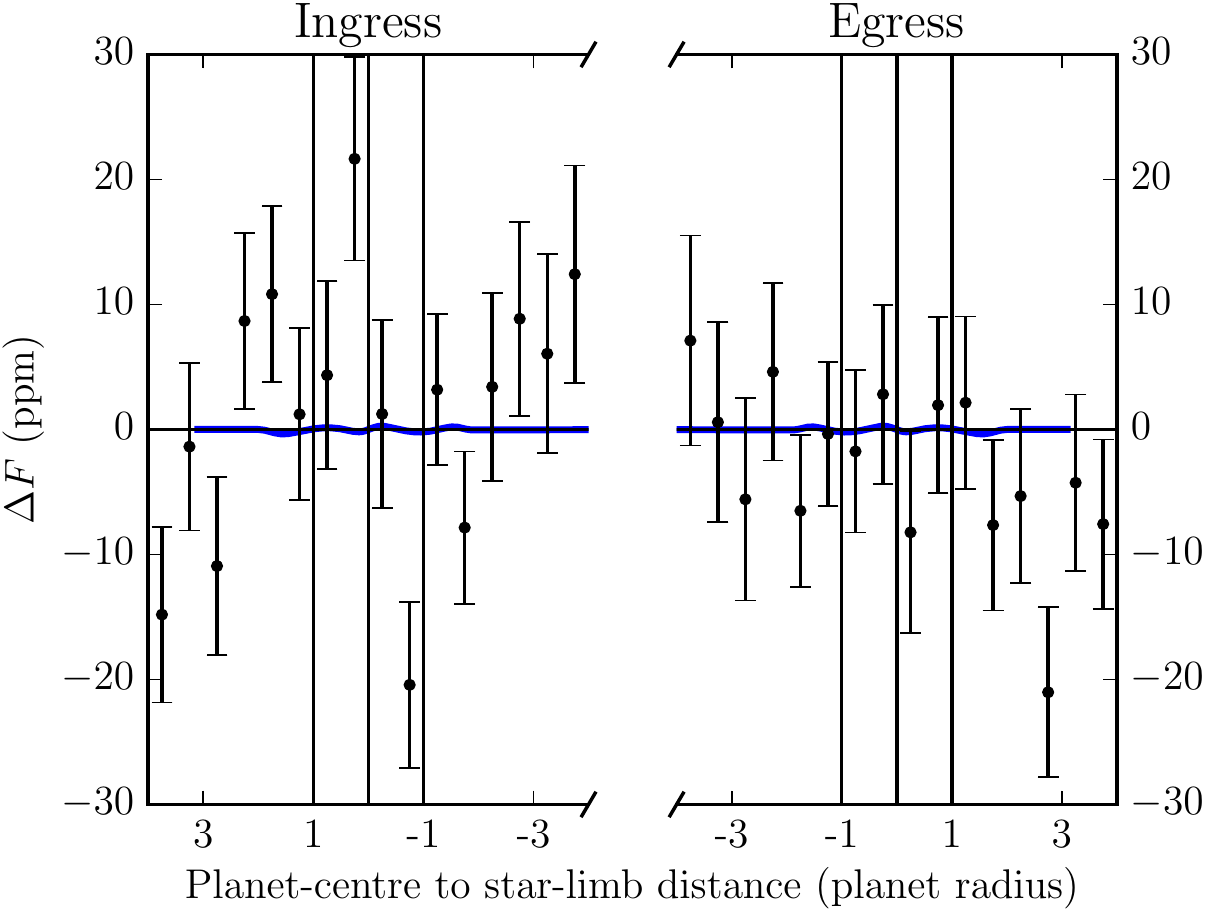}
  \caption{Binned residuals for the sample of large planets on long
    orbits. Contrary to Fig.~\ref{fig:stack_wings}, the $x$-axis is the distance from the stellar limb to the planet centre in units of planet radii. The blue line is the theoretical prediction for a Jovian
    planet on a 20-day orbit. It is equivalent to the blue line in
    Fig.~\ref{fig:mod_run} as observed through the 30-minute long
    cadence of \textit{Kepler}. Negative distances are when planets
    are inside of the stellar disc and positive distances are
    outside. The vertical black lines mark when the planet limb and
    centre cross the star limb.\label{fig:stack_edges}}
\end{figure}
Figure~\ref{fig:stack_edges} shows the stacked residuals around
ingress and egress for the sample of 305 large planets with long
orbits. The sample is the same as in Fig.~\ref{fig:stack_wings}, but contrary to Fig.~\ref{fig:stack_wings}, the data in Fig.~\ref{fig:stack_edges} have the $x$-axis in units of planet radii. Additionally, the distance on the $x$-axis is the distance from the stellar limb
to the planet centre. A total of 28\,911 data points are stacked, with an average of
903 per bin. The choice of unit is made
because refraction features around ingress and egress are expected to
scale with planet size. This is in contrast with the refraction shoulders
that scale with the planet-centre to star-centre distance. Furthermore,
each bin is half a planet radius wide allowing for resolution of
features on shorter timescales. The purpose of this figure is to reveal
any potential features that are similar to those shown in
Fig.~\ref{fig:20d_jovian_param_space}. This would be the only
refraction signal for planets that are not strongly lensing. The
signal is expected to be a few tenths of ppm at \textit{Kepler} long
cadence and the precision of the stacked residuals is approximately
7~ppm. Thus, any refraction features are not expected to be
detectable in this case.

\begin{figure}
  \centering
  \includegraphics[width=\hsize]{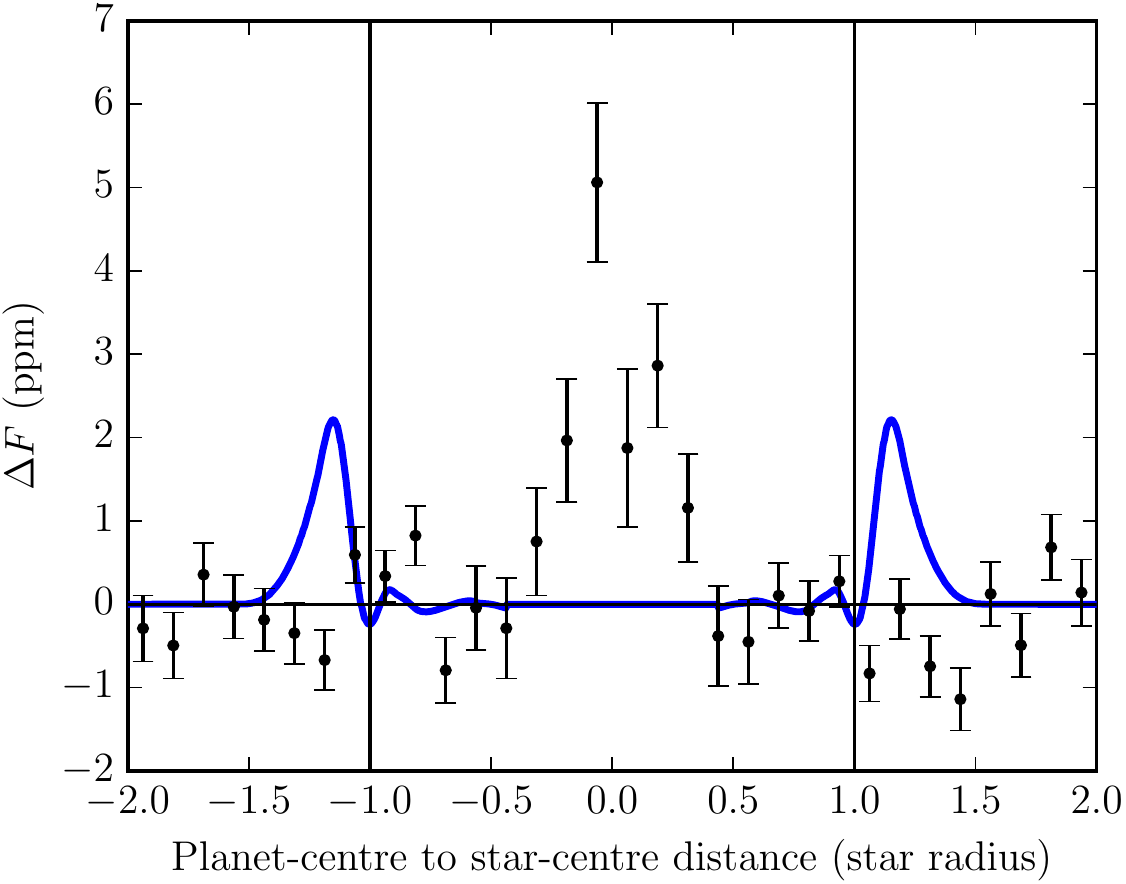}
  \caption{Same as Fig.~\ref{fig:stack_wings}, but for all 2394 KOIs
    that were fitted. The blue line was computed for a 1-year Jovian
    and is not representative for the stacked planets. It is not the
    expected signal for the sample and is included for reference
    only.\label{fig:stack_all}}
\end{figure}
A blind search for any systematic residuals at a high precision in the
sample of all 2394 fitted KOIs is made by stacking the residuals at a
bin size of 0.125 star radii, shown in Fig.~\ref{fig:stack_all}. The
stack contains a total of 2.7~million long-cadence data points with an
average of 84~thousand per bin. The precision during the transit
ranges from 0.3--1~ppm. The residual excess centred on mid-transit possibly stems from inaccurate TTV fitting. The three-parameter TTV model 
(Sect.~\ref{sec:reduction}) lowered the reduced $\chi^2$ value from 
1.481 for $5.34\times{}10^{6}$~DOF to 1.469 for $5.46\times{}10^{6}$~DOF (more DOF with the TTV model because more transits passed all tests). The distributions of reduced $\chi^{2}$ values both with and without TTV modelling are skewed towards large values and have medians of 1.237 and 1.239, respectively. A TTV model with more freedom which allowed each transit to shift slightly in time and another model allowing each transit to shift with respect to the previous one were tested. Both were discarded because they introduced artefacts around ingress and egress. 
It is clear from Fig.~\ref{fig:stack_all} that no
systematic residuals are observed out of transit. Any feature at a few
ppm, such as signs of rings or moons, would be obvious.

No significant differences to the presented results are observed when
dividing the KOIs into other subsets. This includes separating the
sample into narrow ranges of star--planet distances and planet
sizes. Effects of other fitting parameters and selection criteria are
also studied.

\section{Discussion}\label{sec:discussion}
\subsection{Connection with physical properties}
The refraction strength is very sensitive to the atmospheric
scale height, composition, and density, potentially allowing for
constraints on the atmosphere using solely photometric data.
However, parameters such as the
planet radius and orbital distance are more strongly
constrained by other observational methods than atmospheric
lensing.

Refraction also allows for constraints on the presence of clouds or hazes
in the atmosphere. No refraction can be observed in atmospheric layers
that are opaque due to clouds, hazes, or other obscuring weather
conditions, implying that any refraction features can be taken as an
indication of relatively clear skies. Searching for refraction
residuals using photometry alone can therefore yield
suitable targets for follow-up observations, as previously
pointed out by \citet{misra14b}. An alternative approach to detect refraction is to measure molecular feature
wing steepness and relative depths of absorption features in
near-infrared transmission spectra~\citep{benneke13}.

Refraction shoulders, which are only possible in the strong lensing
regime, are likely to be easier to observe than any ingress, egress,
or in-transit residuals. Refraction will be very challenging to detect
for weakly lensing systems because the timescales are much shorter. We
note that the refraction residuals are usually compensated for when
adjusting a transit model to the data. These residuals can also be
confused with stellar oscillations, starspots, and other minor effects
such as rings and moons~\citep{hippke15, kipping15, heller17}. An
additional source of error is the accuracy of the stellar surface
brightness profile. The nature of the refraction shoulders, being out
of transit, facilitates the interpretation of the refraction signal.

There is no simple relation between most physical parameters and
refraction strength. It is a trade-off that depends on many
parameters, as is shown in Sect.~\ref{sec:refraction_shoulders}. The
only predictor for refraction that is robust is if the system is
strongly lensing, but this does not immediately translate to signal
strength. The exact position in the $BC$-plane is not simply
correlated with refraction signal strength, although it does give some
information on qualitative features (e.g.\ the overall shape of the
refraction residuals), but contains little quantitative information.
It is worth
emphasising that refraction can set the effective radius only if the
planet is strongly lensing, which is important for transmission
spectroscopy and is further discussed in
Sect.~\ref{sec:transmission}. The largest refraction peak amplitudes
for realistic planets orbiting Sun-like stars are on the order of
10~ppm, see Sect.~\ref{sec:mod_results} and \citet{misra14b}, but
amplitudes as large as 100~ppm have been reported for hypothetical
fiducial planets~\citep{hui02, sidis10}. The Earth and the best-case
Jovian can be used as examples of planets with high refraction signal
strength given their respective orbital distance and radius. The exact signal strength is sensitive to a
set of parameters and only seems to  be strong for certain
combinations, which means that refraction will be an important effect
for a small number of planets that happen to be favourable for
lensing.

\subsection{Effects on transmission spectroscopy}\label{sec:transmission}
Refraction has attracted much attention as it could increase the
effective radius of a planet and set a `surface' above the deepest
layers of the atmosphere, which has critical consequences for
transmission spectroscopy~\citep{garcia_munoz12, betremieux13,
  betremieux14, betremieux16, betremieux17}. This effect is more
pronounced for in-transit transmission spectra and it was emphasised
by \citet{misra14} that deeper layers can be probed out of transit in
such cases. There is a deep connection between refraction shoulders,
which are observed out of transit, and the maximum depth reached by
in-transit transmission spectroscopy. The reason the deepest layers
are obscured in cases where refraction sets an effective surface is
that any light passing through the deep layers are deflected at a
large enough angle such that no light from these regions reaches the
observer while the planet is in transit. This also means that the
light that penetrates deeply into the atmosphere can be observed near
transit  because of the large deflection angles, which was first
pointed out by \citet{sidis10} (their Figs.~2 and~3) and then also by
\citet{misra14} (their Fig.~9). The photon paths that pass the deep
layers and consequently are deflected at large angles also constitute
the refraction shoulders. Combining the above arguments means that
weakly lensing planets cannot have an effective surface set by
refraction because they are not capable of deflecting light at large
enough angles. The scenario is more complicated if the planet is
strongly lensing because some parts of the atmosphere become
inaccessible during parts of or throughout the transit. It is
therefore not necessary for atmospheric layers to be obscured
throughout the entire transit, but that the effective exposure time is
altitude dependent. Obscured deep layers can be recovered to some
extent by observing the out-of-transit spectra in the refraction
shoulders. The signal is weaker out of transit because of the geometry
that effectively only exposes a narrow slit of the atmosphere on the
outer side of the planet to starlight. This means that refraction can
have significant effects on transmission spectra for strongly lensing
planets even if refraction shoulders cannot be detected because of
insufficient photometric precision. In the most extreme cases, the
deepest layers may not even be observable out of transit because
photons beyond a certain depth are effectively trapped by the
atmosphere.  This possibility is clearly conceptualised by
\citet{sidis10} who discuss the trapping of photons in transparent planets
and the idea of a `lower boundary' defined by \citet{betremieux15}.

Refraction is only weakly dependent on wavelength and is unlikely to
introduce any artefacts that can be interpreted as spectral features
in transmission spectra. The wavelength dependence of refraction is
incorporated in the model through $R_0$ and the refractive coefficient $\alpha$ (Appendix~\ref{sec:cauchy_coefficients}). Only the
latter is inherent to refraction, whereas the former is a radius set by
absorption due to other processes. The inherent wavelength
dependence of refraction can be neglected for practically all
realistic cases. However, it is possible that refraction becomes
significant at wavelengths where the atmosphere is optically thin but
very weak close to strong absorption features or towards
shorter wavelengths where Rayleigh scattering becomes important. In
this case, any effects of refraction will be restricted to the
low-opacity regimes. The general guideline is that refraction does
not obscure layers of the atmosphere if the planet is weakly lensing.
Weak lensing could be restricted to certain wavelengths if the opacity is wavelength dependent or it could be restricted to some transits because of transient clouds or hazes.

\subsection{Host star}
Refraction features depend on host star parameters. Smaller stars
result in larger relative amplitudes for all features because all
normalised quantities scale with the ratio of projected planet area to
star area. An extreme case of a peak amplitude of 950~ppm was reported
by \citet{misra14b} for a 200~K Saturn analogue around an M9V
dwarf. The downside is that the luminosity for small dwarf stars is
much lower and orbits with the same incident flux as for Sun-like
stars are closer-in, which means that the transit duration is shorter
but repeats more often. The change in S/N of any refraction signal is
much less than the peak amplitude might suggest when all parameters
are taken into account, as shown in
Sect.~\ref{sec:red_dwarf}. \citet{misra14} and \citet{betremieux14}
reported that deeper layers of planet atmospheres can be probed for
analogous planets around cooler stars. This means that transmission
spectra can contain information from different parts of the atmosphere
although the S/N of photometric residuals is expected to be
approximately the same for different host stars
(Sect.~\ref{sec:red_dwarf}). It was also shown that only three out of
the five strongly lensing planets around Sun-like hosts are strongly
lensing if orbiting an M4V dwarf. Going from strongly to weakly
lensing shows that deeper layers of atmospheres can be probed around
dwarf stars than around Sun-like stars because weakly lensing planets
are guaranteed to have their entire atmospheres exposed throughout
transits, implying that layers that are inaccessible around a Sun-like
star are exposed around red dwarfs.

\subsection{Alternative refraction scenarios}
It is possible for the atmosphere of an exoplanet to lens a background
star, analogously to gravitational microlensing. The refraction model
predicts infinite amplification for background lensing, which clearly
is an artefact of an idealisation. A more realistic guess would be an
amplification of a factor of a few, based on observed gravitational microlensing
light curves. The main difficulty becomes the occurrence rate. The cross
section is a few AU for gravitational lensing compared to a planet
diameter for atmospheric lensing. The brightness increase does not
have to be mixed with the host star light if the lensing planet can be
directly imaged, since the lensed images of the background star will be
in the atmosphere of the exoplanet. Importantly, any lensed light
would have been transmitted through the atmosphere, thereby having its
spectral imprint.

Another consequence of atmospheric lensing is that it is possible for
 planets with a near-transiting geometry to increase the flux without ever occulting
the host star. This is just the effect of having a system that would
have refraction shoulders if it was transiting, but instead showing a
single flux excess peak when it passes the star without entering the
stellar disc.

\subsection{\textit{Kepler} observations}
No refraction signal is observed in the \textit{Kepler} data when
transits are stacked. The achieved precision is not high enough to
constrain refraction in the sample of large planets in long orbits,
which is the population where refraction is expected to be
significant. It is still possible that refraction is relatively
strong for a few targets, but these get suppressed when stacked with the rest of the sample. No attempt was made to constrain refraction in
individual sources because computing refraction light curves is
computationally intensive and the observational precision is clearly
lower than predicted refraction shoulder amplitudes.

A lack of refraction signatures in the stack of all fitted planets is
expected because the transit method is generally biased towards
detecting planets on short orbital periods. The short planet-to-star distance is
disadvantageous for refraction because it requires atmospheres to
deflect light at larger angles. The only systematic residuals that
can be seen in the stack of all planets is the central excess with a
peak amplitude of 5~ppm. 

\section{Conclusions}\label{sec:conclusions}
We have studied the effects of atmospheric refraction on transit light
curves and searched for refraction signatures in data from the
\textit{Kepler} mission. Our main conclusions are the following:
\begin{enumerate}
\item Refraction residuals in photometric data can constrain physical
  properties of exoplanet atmospheres, such as density, scale height,
  and to some extent composition.
\item Out-of-transit refraction shoulders are the most robust and
  detectable residuals induced by atmospheric lensing and can reach
  peak residual excess amplitudes of approximately 10 ppm. More
  probable amplitudes are a few ppm or less for Jovians and sub-ppm
  for super-Earths. In-transit, ingress, and egress features are
  probably very challenging to detect because of short timescales and
  degeneracy with other transit-model parameters.
\item The parameter space is highly complex and depends critically on
  a number of the parameters parametrising a planetary
  system. Refraction will prove important for some targets whereas
  most planets will show very weak refraction signals.
\item Effects of refraction on transmission spectroscopy are deeply
  connected with refraction shoulders. Refraction does not set the
  effective radius if the planet is weakly lensing. The impact of
  refraction on transmission spectra for strongly lensing planets can
  vary greatly.
\item The detectability of refraction signals for planets orbiting red
  dwarfs cannot be assessed by the sole increase in peak residual
  amplitude. A combination of shorter timescale, shorter orbits,
  shorter star--planet distance, and host star properties results in
  similar values of S/N for planets orbiting Sun-like hosts. However,
  the altitudes probed can be different even though the S/N is
  approximately the same.
\item Refraction residuals cannot be observed in stacked
  \textit{Kepler} light curves because of insufficient photometric
  precision. The only systematic deviation across the entire
  \textit{Kepler} sample is a central excess of a few ppm.
\end{enumerate}

While the present paper was in review, we learnt of a similar study by
\citet{dalba17}. It also explores the dependence of peak residual
excess amplitude on atmospheric parameters and shows comparable
results.

\begin{acknowledgements}
  We thank the anonymous referee for the helpful comments that
  improved the manuscript. B.-O.D. acknowledges support from the Swiss
  National Science Foundation in the form of a SNSF Professorship
  (PP00P2-163967).  This paper includes data collected by the
  \textit{Kepler} mission. Funding for the \textit{Kepler} mission is
  provided by the NASA Science Mission directorate. This research has
  made use of the NASA Exoplanet Archive, which is operated by the
  California Institute of Technology, under contract with the National
  Aeronautics and Space Administration under the Exoplanet Exploration
  Program. This research made use of Astropy, a community-developed
  core Python package for Astronomy~\citep{astropy13}. This research
  has made use of NASA's Astrophysics Data System Bibliographic
  Services.
\end{acknowledgements}

\bibliographystyle{aa} 
\bibliography{references} 

\begin{appendix} 
\section{Cauchy's equation}\label{sec:cauchy_coefficients}
\begin{table}
  \centering
  \caption{\label{tab:cauchy_coefs} Cauchy's equation coefficients.}
  \begin{tabular}{l c c}\hline\hline
    Gas composition & $A_1 \times 10^{-5}$ & $B_1 \times 10^{5}$ \\
    &&(\AA{}$^2$)\\\hline
    Hydrogen &            $13.60$ & \hphantom{0}$7.7$ \\
    Helium   & \hphantom{0}$3.48$ & \hphantom{0}$2.3$ \\
    Argon    &            $27.92$ & \hphantom{0}$5.6$ \\
    Nitrogen &            $29.19$ & \hphantom{0}$7.7$ \\
    Oxygen   &            $26.63$ & \hphantom{0}$5.1$ \\
    Methane  &            $42.60$ &            $14.4$ \\
    Ethane   &            $73.65$ & \hphantom{0}$9.1$ \\
    Air      &            $28.79$ & \hphantom{0}$5.7$ \\\hline
  \end{tabular}
  \tablefoot{Values are given for gases at a temperature of
    $0^\circ{}$~C and pressure of $1$~atm~\citep{born99,
      griffiths99}.}
\end{table}
Cauchy's equation is used to compute the refractive coefficient
($\alpha$) for different atmospheric compositions. The refractive
coefficient is not to be confused with the refractive index ($n$),
which are related by $n = 1 + \alpha \rho$, where $\rho$ is the
density. The refractive coefficient is given by
$\alpha \rho = A_1 (1 + B_1/\lambda^2)$, where $A_1$ is the
coefficient of refraction, $B_1$ is the coefficient of dispersion, and
$\lambda$ is the wavelength. Coefficients for some common gases are
provided in Table~\ref{tab:cauchy_coefs}. The refractive coefficients
are calculated at a temperature of $0^\circ{}$~C and pressure of
$1$~atm, and are assumed to be constant over the relevant temperatures
and pressures.

\end{appendix}
\end{document}